\documentclass[fleqn,usenatbib]{mnras}

\usepackage[T1]{fontenc}

\DeclareRobustCommand{\VAN}[3]{#2}
\let\VANthebibliography\thebibliography
\def\thebibliography{\DeclareRobustCommand{\VAN}[3]{##3}\VANthebibliography}

\usepackage{graphicx}	
\usepackage{amsmath}	
\usepackage{amssymb}	
\DeclareUnicodeCharacter{2212}{-}

\title{Universal Relations For Generic Family Of Neutron Star Equations Of State}

\author[Kamal Krishna Nath, Ritam Mallick and Sagnik Chatterjee]{
Kamal Krishna Nath,$^{1}$\thanks{E-mail: knath@iiserb.ac.in}
Ritam Mallick,$^{2}$\thanks{E-mail: mallick@iiserb.ac.in}
Sagnik Chatterjee$^{3}$\thanks{E-mail: sagnik18@iiserb.ac.in}
\\
$^{1}$Indian Institute of Science Education and Research Bhopal\\
$^{2}$Indian Institute of Science Education and Research Bhopal\\
$^{3}$Indian Institute of Science Education and Research Bhopal
}

\date{Accepted XXX. Received YYY; in original form ZZZ}

\pubyear{2023}

\begin{document}
\label{firstpage}
\pagerange{\pageref{firstpage}--\pageref{lastpage}}
\maketitle

\begin{abstract}
Universal relations are important in testing many theories of physics. In the case of general relativity, we have the celebrated no-hair theorem for black holes. Unfortunately, the other compact stars, like neutron stars and white dwarfs, do not have such universal relation. 
However, neutron stars (and quark stars) have recently been found to follow certain universality, the I-Love-Q relations. These relations can provide a greater understanding of the structural and macro properties of compact astrophysical objects with knowledge of any one of the observables. The reason behind this is the lack of sensitivity to the relations with the equation of state of matter. In our present work, we have investigated the consistency of universal relations for a generic family of equations of state, which follows all the recent astrophysical constraints. Although the spread in the EoS is significant the universal nature of the trio holds relatively well up to a certain tolerance limit. The deviation from universality is seen to cross the tolerance limit with EoS, which is characteristically different from the original set. 
\end{abstract}

\begin{keywords}
Moment of Inertia -- Quadrupole moment -- Tidal love number -- Dense matter -- Equation of state

\end{keywords}



\section{Introduction}

Neutron stars (NSs) have proven to be excellent laboratories for testing various laws of physics, starting from general relativity to nuclear and particle physics. They are a good candidate to test general relativity as the gravitational field in and around them is quite strong. They have proved to be a good source which emits gravitational waves (GW) along with electromagnetic waves (EW). Simultaneous detection of GW and EW by GW170817 has paved the way towards multimessenger astronomy \citep{Abbott,Abbott1}. NSs are quite massive (around $1.4-2.1$ solar mass) but are very compact (radius around $10-15$ km), which makes them one of the densest objects in the universe (with density at the cores can be as high as $8-10$ times nuclear saturation density). This makes them a vital candidate to test the theory of strongly interacting matter at high density, which is still one of the least understood states of matter. The theory of strongly interacting matter (quantum chromodynamics) predicts a phase transition from a confined hadronic state to a deconfined quark state at intermediate densities \citep{Shuryak}. However, theoretical calculations and experiments at such densities have proved difficult to perform. Therefore, the only laboratory to test these theories is the NSs.

However, testing theories has been challenging with observations coming from NSs. This is because the emission (GW and EW) comes from the star's surface, and the core is completely hidden where the physics of high-density matter lies. To understand the physics of the star's interiors, one has to model them carefully from the interior to the surface and then match them with observational signatures that originate from them. Previously, there has been only electromagnetic observation from pulsars (rotating NSs) like radio pulses, X-rays and gamma rays. However, the picture changed drastically after the observation of GW from GW170817. We found another probe to examine the NSs.

Simultaneous measurement of a few important characteristics feature of NSs helps in understanding the matter properties (basically the equation of state (EoS) of matter) of NS interior in a better way. The mass and radius measurement would constrain the EoS severely. Although the precise measurement of mass has been accomplished to a certain degree of accuracy, the radius measurement has proved to be a challenge for a long. Recently, the NICER mission has been launched to measure the mass-radius of pulsars to a high degree of accuracy and has produced significant results like the measurement of pulsars PSR J0030+0451 and PSR J0740+6620 \citep{riley,miller,riley1,miller1}. This measurement has significantly constrained the EoS to a certain degree, but still, ambiguities remain on whether the star's core has confined or deconfined matter.

On the other hand, GW170817 has also contributed towards constraining the EoS of matter. The detection of the GW in binaries during its late inspiral helps deduce the star's tidal deformability. The tidal deformability is related to the NS structure \citep{hinderer,damour,binnington}, which in turn depends on the EoS of matter \citep{flanagan,hinderer,Kyutoku,lackey,baiotti,bernuzzi}. The detection of GW170817 has also helped in constraining the EoS by constraining the binary tidal deformability ($\widetilde{\Lambda} \le 720 $) \citep{altiparmak}. The NICER and GW170817 results have greatly helped in constraining the EoS; however, the ambiguity remains. This is because the exterior gravitational field depends not only on the mass but also on the moment of inertia and quadrupole moment (the leading order for the generation of GW). The degeneracies in the NS spin and quadrupole moment hinder their detection separately from GW in binaries.

The physical parameters that characterise the NS should depend on the EoS. However, it has been shown that some combination of the physical parameters does not depend on the details of the EoS of matter and follows some universal relations \citep{yagi,rezzolla,debojoti}. Other than the mass (M) and radius (R), one can calculate the moment of inertia (I, characterizing the spin of the NS), the quadrupole moment (Q, the deviation from spherical symmetry) and the tidal deformability (or the love number ($\lambda$) which characterizes the deformability of the star in an external tidal field). Although intuitively, they should be dependent on the EoS, it was shown that they are not (very weak dependence). It has also been argued that NS being a compact object, should follow a similar relation to the no-hair theorem of black holes (BH) \citep{robinson,hawking}. Although NSs do not have such a relation, they approach the limit of universality (independent of the EoS) for these trio. 

Universal relations are important because if we know one parameter, we can calculate the others knowing the relation. In the case of NSs, if we can somehow measure I, we can calculate the love number and Q without actually measuring them. This is very helpful because both are difficult to measure for binaries which are at a considerable distance. For coalescing binaries where one can measure the love number quite accurately, one can deduce the Q and, thereby, some information about the spins of the coalescing binary NSs \citep{Abbott1}. The universal relation can also be used to test the theory of general relativity (GR) and the correction of the modified theory of gravities. However, the universal relation has its limitations, and one needs to be careful while calculating individual quantities. It was shown in \citep{Pani} that if the rotational love number is considered in place of the non-rotating love numbers, the universality gets violated. The first study of the I-Q relation for rapidly rotating NSs was done by Doneva \citep{doneva}. They found that the I-Q relation is broken and becomes more EOS-dependent for NSs with a fixed frequency. However, it was soon found by Pappas \& Apostolatos \citep{pappas}, and Chakrabarti et al. \citep{chakra} that the I-Q relation can remain approximately EOS-insensitive if one chooses suitable dimensionless parameters instead of dimensional quantities.

Until recent years, tests of the I-Love-Q relations have largely been confined to a limited set of tabulated EoS. Expansion of the relation to unexplored regions of EoS parameter space has also been done \citep{benitez}. Sham et al. \citep{Sham} showed that the I-Love-Q relations of low-mass NSs near the minimum mass limit depend more sensitively on the underlying EoS as these stars are composed mainly of softer EoS. Haskell et al. \citep{Haskell} explored the I-Love-Q relations for the magnetized NSs. Different magnetic field geometries provide different I–Q relations. In the case of a more realistic twisted-torus magnetic field configuration, the relation depends significantly on the EoS, thus, losing its universality. Universality is lost for stars with long spin periods, i.e. P $>$ 10s, and strong magnetic fields, i.e. B $>$ $10^{12}$ G. The I-Love-Q relations for super-fluid NSs were studied by Yeung et al. \citep{yeung}. The article studied the extent to which the two-fluid dynamics might affect the robustness of the I-Love-Q relations by using a simple two-component polytropic model and a relativistic mean-field model.

But how robust are this universal relation of the trio for a generic family of EoSs obeying all the astrophysical constraints? Recently, there have been agnostic approaches to constrain the EoS of matter at high densities \citep{Annala,rezzolla,altiparmak,ecker2}. The EoS of matter at low densities are more or less constrained and are known with a certain degree of accuracy \citep{bps}. Also, at high-density, perturbative QCD calculation predicts quite robust results \citep{Kurkela0}. Using these two boundary conditions and the fact that the adiabatic speed of sound ($c_s$) can only lie between zero and one, one can construct numerous EoS \citep{Annala}. The speed of sound is vital in determining the EoS as it gives the relation between pressure (p) and energy density ($\epsilon$), $c_s=\sqrt{\frac{\partial p}{\partial \epsilon}}$. Once the EoS region is defined with the approach, one can then further constrain the EoS with the recent measurement from NICER and GW170817. Although the method does not specifically tell us about microphysics, one gets a very good bound on the overall nature of the EoS. In this work, we construct a collection of EoS and then test the universal relation of the I-love-Q. 

The paper is arranged in the following way: in section II we give the formalism that we have used to calculate the collection of agnostic EoS and the I, love and Q values for the stars. Next, in section III, we give our results about the universal nature of the I-love-Q relation and also the error in the universality. Finally, section IV summarises our results and draws important conclusions.

\section{Formalism}

To check the universal relation for a generic family of EoS, first, we need to construct the agnostic EoS sets as done in the seminal paper \citep{Annala}. The EoS of nuclear/hadronic matter (NM) up to the nuclear saturation density of $n_0\approx 0.16$ fm$^{-3}$ is well-understood \citep{bps}. However, close to this value, the underlying uncertainties in most nuclear matter calculations start to increase rapidly. Nuclear-theory tools such as chiral effective theory (CET) are limited to densities below saturation densities \citep{heb,cet}. Therefore, below saturation density, we choose the EoS, which is allowed at low densities \citep{Hebeler}. Interestingly there are also small uncertainties even at such sub-saturation densities. However, this does not contribute much to the uncertainty in the NS structure. Conversely, perturbative QCD (pQCD) becomes reliable only at much higher densities \citep{lat,fraga,Kurkela0}. No first-principles calculations are applicable at intermediate densities encountered inside the NS cores. However, it is expected that whatever EoS the intermediate densities have, they should converge to the pQCD limit at asymptotic densities.

It is important to obtain robust information on the properties of NS matter at core densities. In particular, there is a requirement that the EoS must reach its known low and high-density limits while behaving in a thermodynamically consistent manner \citep{Annala}. In this paper, we take a similar approach to construct a family of EoSs that interpolate between a CET EoS below saturation density and pQCD result at high densities. In this respect, the adiabatic speed of sound $c_s$ has been used to obtain the EoS by randomising it arbitrarily in the intermediate densities. The $c_s$ gives the pressure energy relation and helps determine the EoS. Also, $c_s$ is bounded in the limit $0 \le c_s \le 1$.
We randomized the speed of sound in $5$ different pieces as a function of chemical-potential segments in the density range $1.1\,n_0 < n \lesssim 40\,n_0$. We used the sound-speed parametrization method introduced in \citep{Annala}. This method uses the sound speed as a function of the chemical potential $\mu$ as a starting point to construct thermodynamic quantities. The piecewise linear segments for the sound speed of the following form:

\begin{equation}
 c_s^2(\mu)=\frac{\left(\mu _{i+1}-\mu \right) c_{s,i}^2+\left(\mu -\mu
   _i\right) c_{s,i+1}^2{}}{\mu _{i+1}-\mu _i}\,,
   \label{cs2}
\end{equation}

where $\mu_i$ and $c_{s,i}^2$ are parameters of the $i$-th segment in the
range $\mu_i\le \mu \le \mu_{i+1}$. 
As the final step in our procedure, we keep solutions whose pressure, density, and sound speed at $\mu_{i}=2.6\,{\rm GeV}$ are consistent with the parametrized perturbative QCD result for cold quark matter in beta equilibrium \citep{Kurkela1}.
%
%

In the next step, we have imposed the observational constraints. The first step is to compute the mass-radius relation for each EOS by numerically solving the Tolman-Oppenheimer-Volkoff (TOV) equations. Then we impose the mass measurements of J0348+0432~\citep{Antoniadis} ($M=2.01\pm 0.04~M_\odot$) and of J0740+6620~\citep{cromartie} ($M=2.08 \pm 0.07 M_\odot$) by rejecting EOSs that have a maximum mass $M_{_{\rm TOV}}<2.0\,M_\odot$. In addition, we impose the constraints subject to radius measurements by NICER of J0740+6620~\citep{riley1,miller1} and of J0740+6620~\citep{riley,miller} by rejecting EOSs with $R<10.75\,{\rm km}$ at $M=2.0\, M_\odot$ and $R<10.8\,{\rm km}$ at $M=1.1\, M_\odot$, respectively. Recently, there has been an observation of a massive pulsar PSR J0952−0607 which gives a maximum mass bound in the range $2.18 - 2.52$ \citep{romani}. In this article, we have also highlighted the latest maximum mass constraint and its effect on the universal relation.

Once we have constructed the generic family of EoS and have obtained their mass-radius sequence, we calculate the I-love-Q trio for each EoS.
In the presence of a companion, an NS (or, as a matter of fact, even a quark star (QS)) is quadrupolar deformed. The quadrupole moment tensor $Q_{ij}$ determines the magnitude of this deformation and can be written as  $Q_{ij} = - \lambda \mathcal{E}_{ij}$, where $\lambda$ is the tidal Love number and $\mathcal{E}_{ij}$ is the quadrupole (gravitoelectric) tidal tensor that characterizes the source of the perturbation. Here we introduce the dimensionless tidal Love number $\bar{\lambda} = \lambda/M^5$, which physically characterizes the tidal deformability of a star in the presence of the companion's tidal field. ${\lambda}$ can also be calculated by treating the tidal effect of the companion star as the perturbation to the isolated (non-rotating) NS. To calculate tidal love numbers of individual stars, we first solve the TOV equations for static NSs and solve the following expression for $k2$ \citep{hinderer}

 \begin{eqnarray}
&& k_2 = \frac{8C^5}{5}\left(1-2C\right)^2
\left[2+2C\left(y-1\right)-y\right]\times\nonumber\\
&&\bigg\{2C\left(6-3 y+3 C(5y-8)\right)\nonumber\\
&&
+4C^3\left[13-11y+C(3 y-2)+2
C^2(1+y)\right] \nonumber\\
&& ~ ~
+3(1-2C)^2\left[2-y+2C(y-1)\right]\log\left(1-2C\right)\bigg\}^{-1},\nonumber
\end{eqnarray}

where $\lambda$ is related to $k_2$ as follows:
\begin{equation}
k_2 = \frac{3}{2}G \lambda R^{-5}.
\end{equation}
where $G$ is the gravitational constant and $R$ is the radius of the star.

Finally, we exploit the detection of GW171817 by LIGO/Virgo to set an upper bound on the tidal deformability for $1.4$ $M_{\odot}$ NS, $\bar{\lambda}_{1.4}<800$ \citep{Abbott,Abbott1}. The EoSs should be able to produce 1.4 $M_{\odot}$ NSs with dimensionless tidal love number $\bar{\lambda}$ (also called the tidal parameter) less than 800. $\bar{\lambda}_{1.4}$ is the tidal parameter specifically for the 1.4 $M_{\odot}$ NS.\\

In this paper, we have considered both isolated and rotating NSs that are described by their mass $M$, the magnitude of their spin angular momentum $J$ and angular velocity $\Omega$, its (spin-induced) quadrupole moment $Q$ and their moment of inertia $I \equiv J/\Omega$.  We have introduced the dimensionless quantities $\bar{I} \equiv I/M^{3}$ and $\bar{Q} \equiv -Q/(M^{3} \chi^{2})$, where $\chi \equiv J/M^{2}$ is the dimensionless spin parameter. The quantities introduced above have an explicit physical meaning: $I$ determines how fast a body can spin given a fixed $J$; $Q$ specifies the amount of stellar quadrupolar deformation. These quantities are determined by modelling the NSs using RNS code \citep{stergioulas,nozawa}. We set up RNS to use a grid of DMDIV = 151 and DSDIV = 301 (angular times radial grid points) and a tolerance in the desired value of the parameters of $10^{-4}$.\\

\section{Results}
\subsection{Non-monotonic Sound Speed in NSs}

\begin{figure}
\includegraphics[width=\columnwidth]{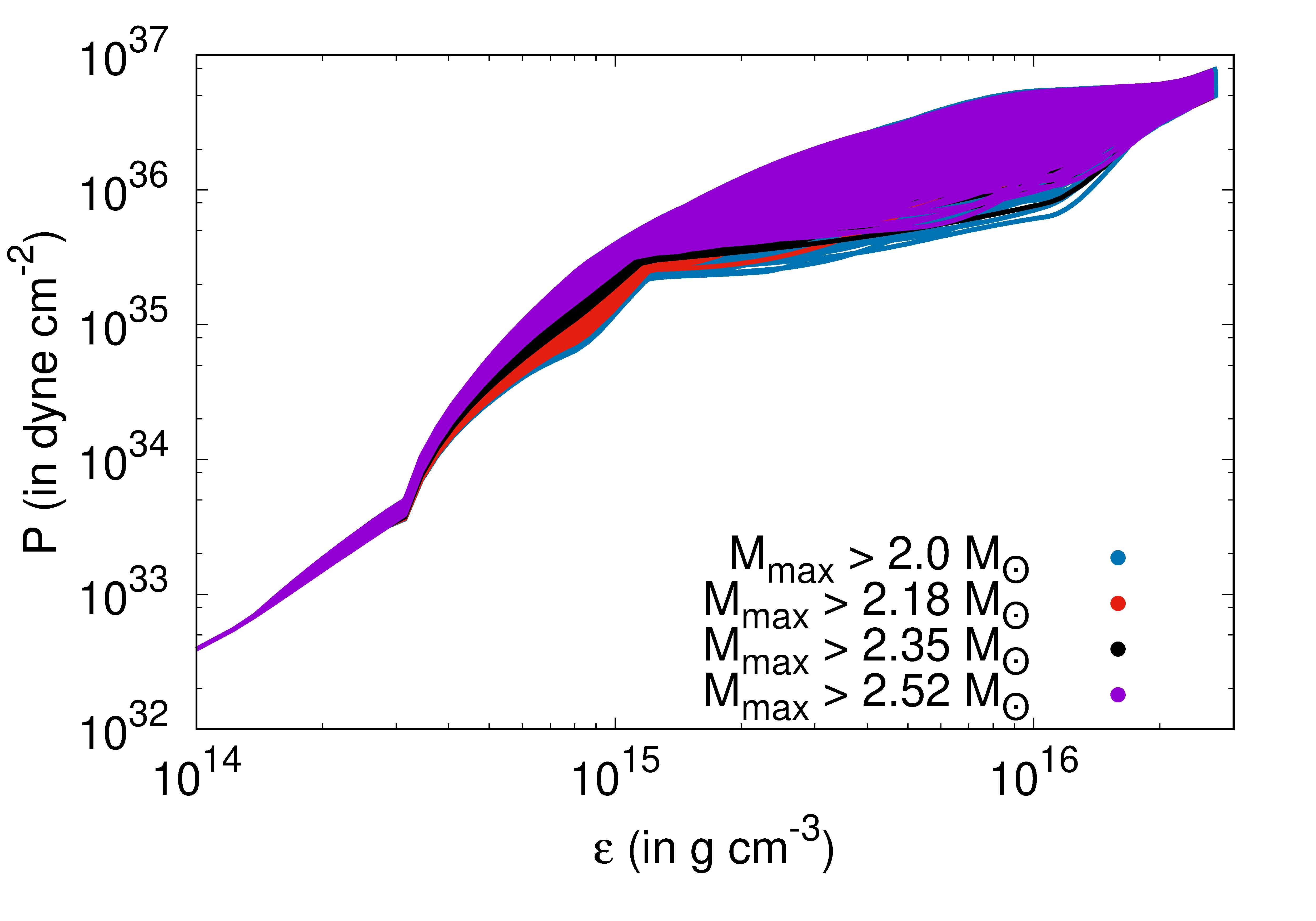}
    \caption{The plot shows the family of 10000 EoSs constructed by parameterizing sound speed that follows all astrophysical observational constraints. The different colour grading signifies EoSs with maximum masses above $4$ different values. The blue colour signifies the EoSs with maximum mass above 2 $M_{\odot}$. The red, black and purple colours denote the EoSs with a maximum mass of 2.18 $M_{\odot}$, 2.35 $M_{\odot}$ and 2.52 $M_{\odot}$ respectively, which form a subset of the EoSs with maximum mass above 2 $M_{\odot}$.}
    \label{f1}
\end{figure}

The pQCD calculations predict that at high density, the speed of sound should reach the conformal limit, which is $\frac{c}{3}$. All the EoS satisfies the given bound at asymptotic densities; however, at intermediate densities, they cross the limit of $\frac{1}{3}$ ($c_s = 1$ in geometrized units). Therefore, there is a peak in the intermediate densities for the sound speed for most of the EoSs, and they become non-monotonous. This is necessary to generate massive stars.
Fig \ref{f1} shows all the $10000$ EoSs constructed using the above formalism. The $p$ vs $\epsilon$ curve is plotted in the logarithmic scale to clearly depict the entire range of densities (starting below saturation till beyond the pQCD limit). All the $10000$ EoS (marked in blue) are able to produce NSs with a maximum mass of more than $2$ $M_{\odot}$ and satisfy the NICER/GW170817 constraints. The red, black and purple colour signifies EoS (regions) whose maximum mass are equal and above 2.18 $M_{\odot}$, 2.35 $M_{\odot}$ and 2.52 $M_{\odot}$ respectively; however, they still obey the NICER/GW170817 constraints. As expected, as the maximum mass increases, the region shrinks. Even for a maximum mass of $2.52$ $M_{\odot}$, there are about 3000 EoSs which satisfy the observational constraints.

\begin{figure}
	\includegraphics[width=\columnwidth]{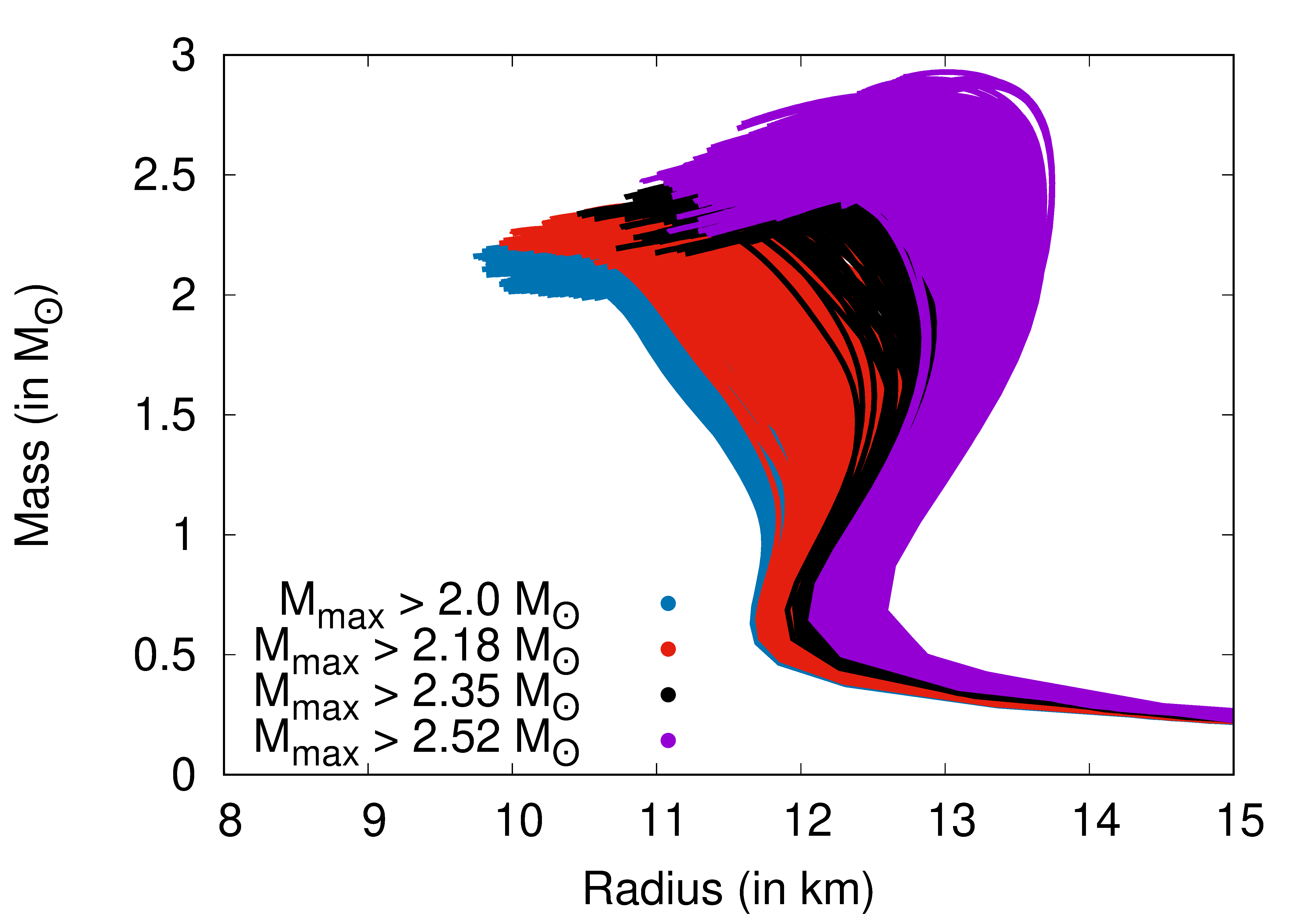}
    \caption{M-R curve for the family of EoSs. Similar to the fig \ref{f1}, the M-R curve of EoSs with maximum mass above 2 $M_{\odot}$ are shown in blue. The EoSs with 3 other maximum mass bounds is shown which form a fraction of the generic family. The nomenclature for the colour grading remains the same.}
    \label{f2}
\end{figure}

Once we construct the EoSs, we solve the TOV equation to generate the mass-radius sequence of the NSs (fig \ref{f2}). Although we call them NSs, they can even be QSs. However, as the EoS curves are smoothly connected, the phase transition from the confined state to the deconfined state cannot happen with first-order phase transition as there is no density/chemical potential jump. Similar to that of fig \ref{f1}, different colour-shaded regions define different maximum mass bounds and have the same nomenclature. 

\begin{figure}
	\includegraphics[width=\columnwidth]{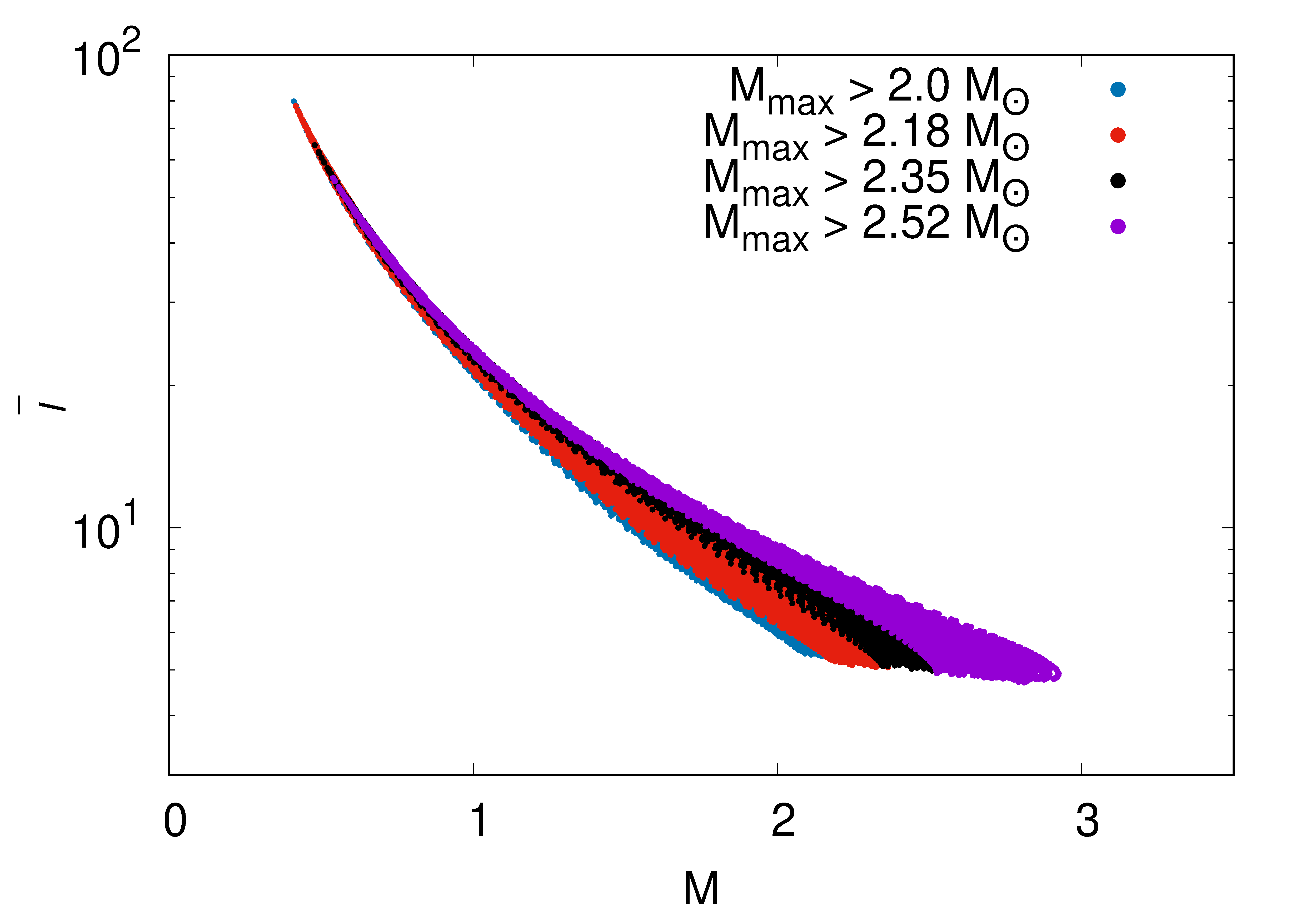}
    \caption{Moment of inertia of all the NSs is plotted as a function of their gravitational mass. The nomenclature for the colour grading remains the same as fig \ref{f1}. The NSs with EoSs of different maximum mass bounds seem to form different regions with a common area similar to the M-R curve.}
    \label{f3}
\end{figure}

Fig \ref{f3} shows how the moment of inertia changes with respect to the gravitational mass. Every point denotes an NS with a given moment of inertia and mass. The stars with lower mass have a larger moment of inertia because their radius is quite large. $\bar{I}$ has a considerable spread as a function of gravitational mass.

\begin{figure}
	\includegraphics[width=\columnwidth]{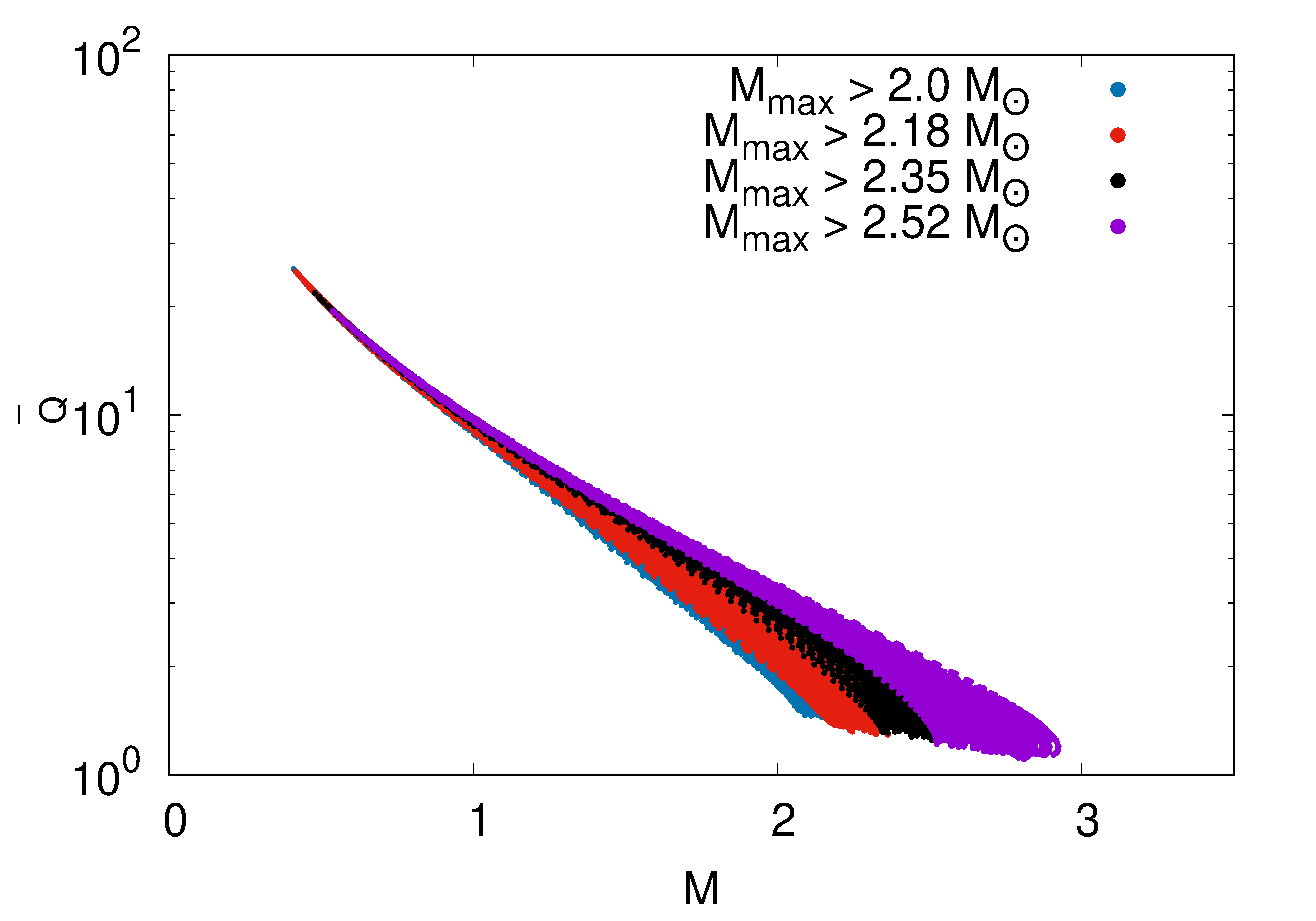}
	\caption{Quadrupole moment of all the NSs is plotted as a function of their gravitational mass. The nomenclature for the colour grading remains the same as fig \ref{f1}. The NSs with EoSs of different maximum mass bounds seem to form different regions with a common area similar to the M-R curve. This difference in the values can be related to the stiffness of the EoSs.}
	\label{f4}
\end{figure}

In fig \ref{f4}, the quadrupole moment is plotted with respect to the gravitational mass of NS. The quadrupole moment behaves very similarly to that of $\bar{I}$. It has a larger value for smaller stars and decreases as we move towards more massive stars. $\bar{Q}$ smoothly falls as we move to higher masses. Also, there is a significant spread for the quadrupole moment. Both $\bar{I}$ and $\bar{Q}$ depend strongly on the star's rotation. In our calculation, we assumed the star's rotation frequency to be much below the mass shedding frequency, $\Omega \approx 450$ Hz. For a fixed value of $\Omega$, the low mass NSs have very minimal spread as compared to the higher mass counterparts. This is because the low-mass stars have low central densities which are below the saturation density. At these sub-nuclear densities, the EoS is highly constrained and matter properties are similar, thus, resulting in confined values of quadrupole moment and moment of inertia. 
 
\begin{figure}
	\includegraphics[width=\columnwidth]{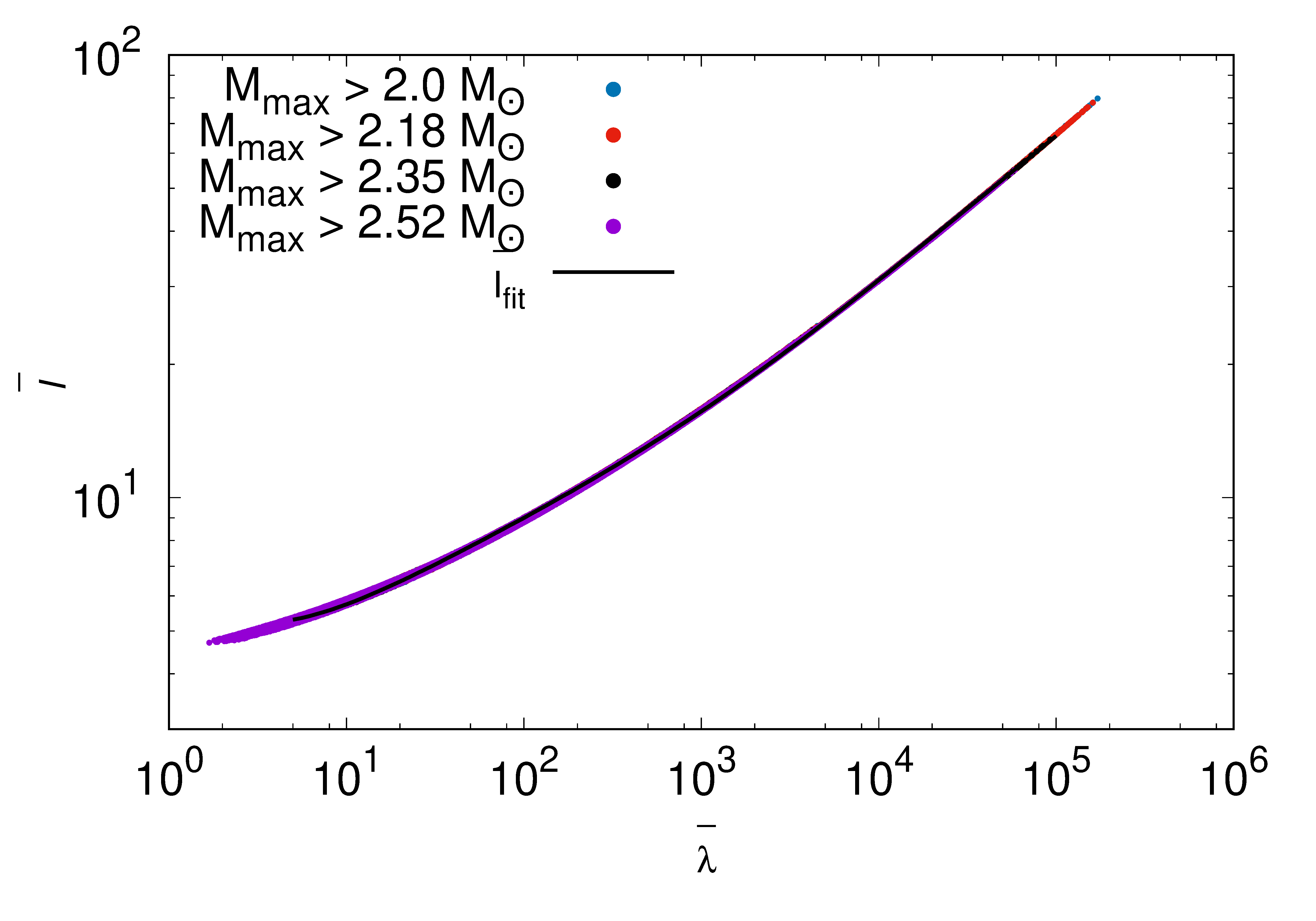}
	\caption{Moment of inertia of the NSs is plotted with respect to their tidal love number. The nomenclature for the colour grading remains the same as fig \ref{f1}. The black line denotes the fitting function $\bar{I}_{fit}$. This function defined in eqn \ref{fit} using the fitting parameters (for $\bar{I}-\bar{\lambda}$ relation) mentioned in table \ref{fit-tab}.}
	\label{f5}
\end{figure}

\begin{figure}
	\includegraphics[width=\columnwidth]{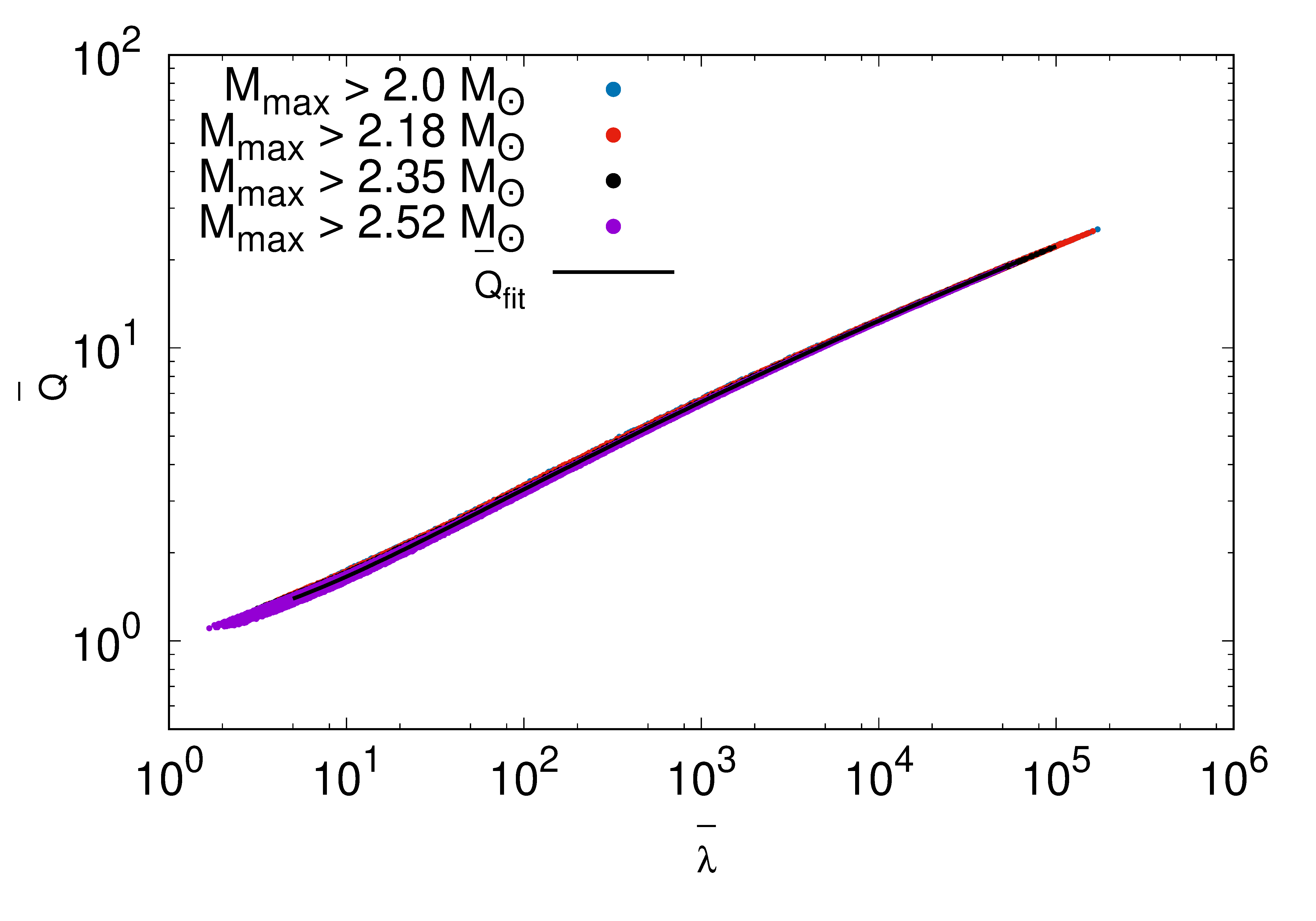}
	\caption{Quadrupole moment of the NSs is plotted with respect to their tidal love number. The nomenclature for the colour grading remains the same as fig \ref{f1}. The black line denotes the fitting function $\bar{Q}_{fit}$. The function is defined in eqn \ref{fit} using the fitting parameters (for $\bar{Q}-\bar{\lambda}$ relation) mentioned in table \ref{fit-tab}.}
	\label{f6}
\end{figure}

Out of $10000$ EoSs we randomly choose $1000$ EoSs each for different maximum mass limits. The number of EoSs, rotational frequency and grid size of the numerics are kept consistent for the EoSs with all the 4 maximum mass bounds to avoid any bias.
Although both $\bar{I}$ and $\bar{Q}$ show some regularity when plotted as a function of gravitational mass, they have quite a spread. However, the picture changes drastically when $I$ is plotted as a function of love number $\bar{\lambda}$ (fig \ref{f5}). They almost follow a curve with minimal spread. The maximum mass constraints have almost no effect on the universality; only the maximum value of $\bar{\lambda}$ changes. EoS with higher maximum mass ends at lower $\bar{\lambda}$ values. A similar nature is seen when we plot $\bar{Q}$ as a function of $\bar{\lambda}$. For $1000$ EoSs each for different maximum mass limits, the universality between the two holds (fig \ref{f6}).

It is useful to fit the universal relation with a polynomial on a log scale. Then measuring one of the trio, one can estimate the other two parameter values without measuring them directly. The polynomial fit is given by  

\begin{equation}
y_i = a_i + b_i \ln x_i + c_i (\ln x_i)^2 + d_i (\ln x_i)^3+ e_i (\ln x_i)^4\,,
\label{fit}
\end{equation}
where the coefficients are summarized in table \ref{fit-tab}. 

\begin{table}
\begin{tabular}{ccccccc}
\hline
\hline
\noalign{\smallskip}
 $y_i$ & $x_i$ &{$a_i$} &{$b_i$}
& {$c_i$} & {$d_i$} & {$e_i$} \\
\hline
\noalign{\smallskip}
 $\bar{Q}$ & $\bar{\lambda}$ & 1.1365 & 0.01691 & 0.5179 & -0.0326 & 0.0194\\
 ${\bar{I}}$ & $\bar{\lambda}$ & 5.9833 & -3.366 & 4.259 & -1.3574 & 0.2235\\
\noalign{\smallskip}
\hline
\hline
\end{tabular}
\caption{Estimated numerical coefficients for the fitting formulas of the NS and QS I-Love and Q-Love relations.}
\label{fit-tab}
\end{table}

\begin{figure}
	\includegraphics[width=\columnwidth]{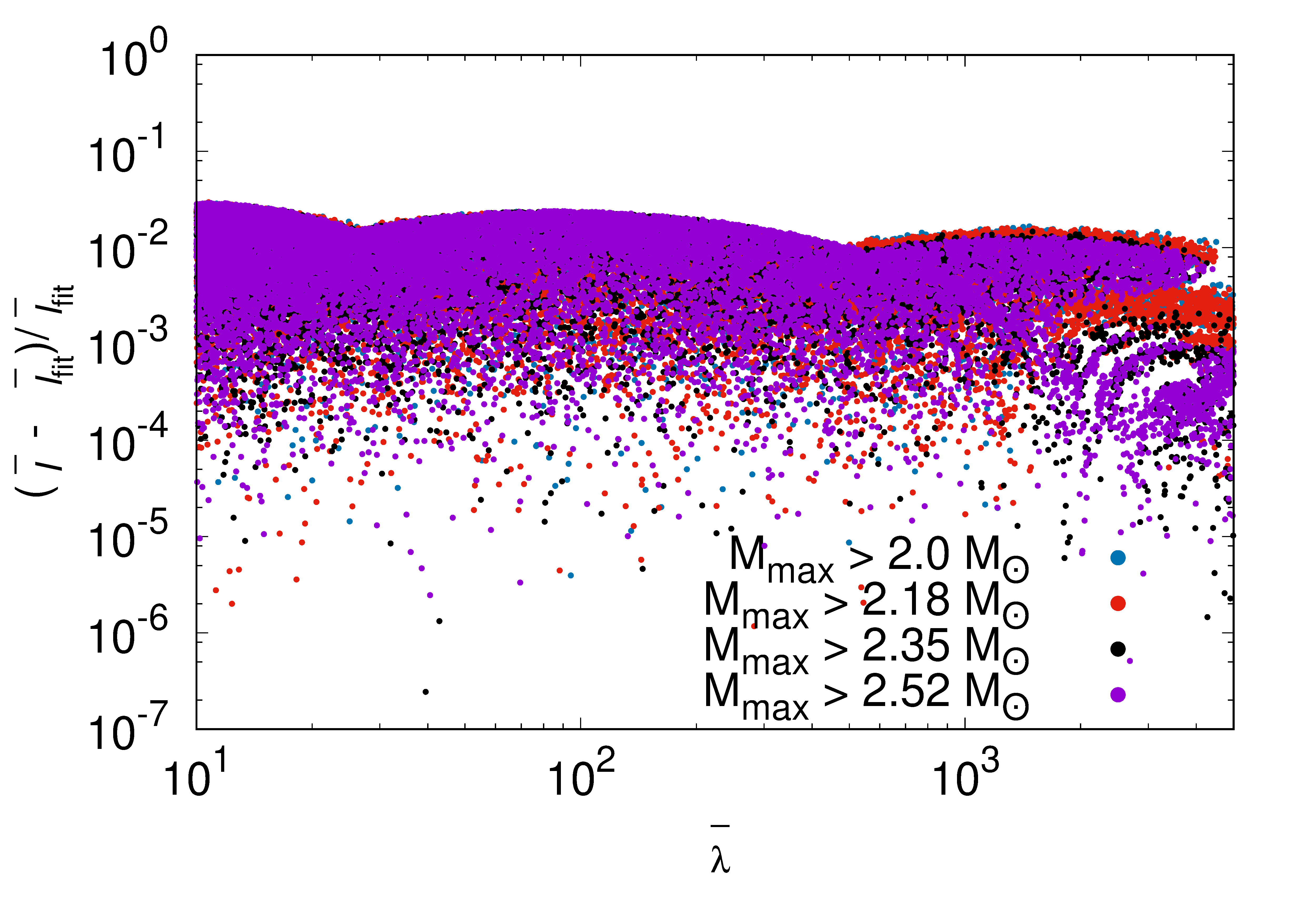}
	\caption{The fractional error in the moment of inertia of NSs from the fitting function $\bar{I}_{fit}$ is shown. The nomenclature for the colour grading remains the same as fig \ref{f1}. The different maximum mass bounds don't separate out into different regions.}
	\label{f7}
\end{figure}

\begin{figure}
	\includegraphics[width=\columnwidth]{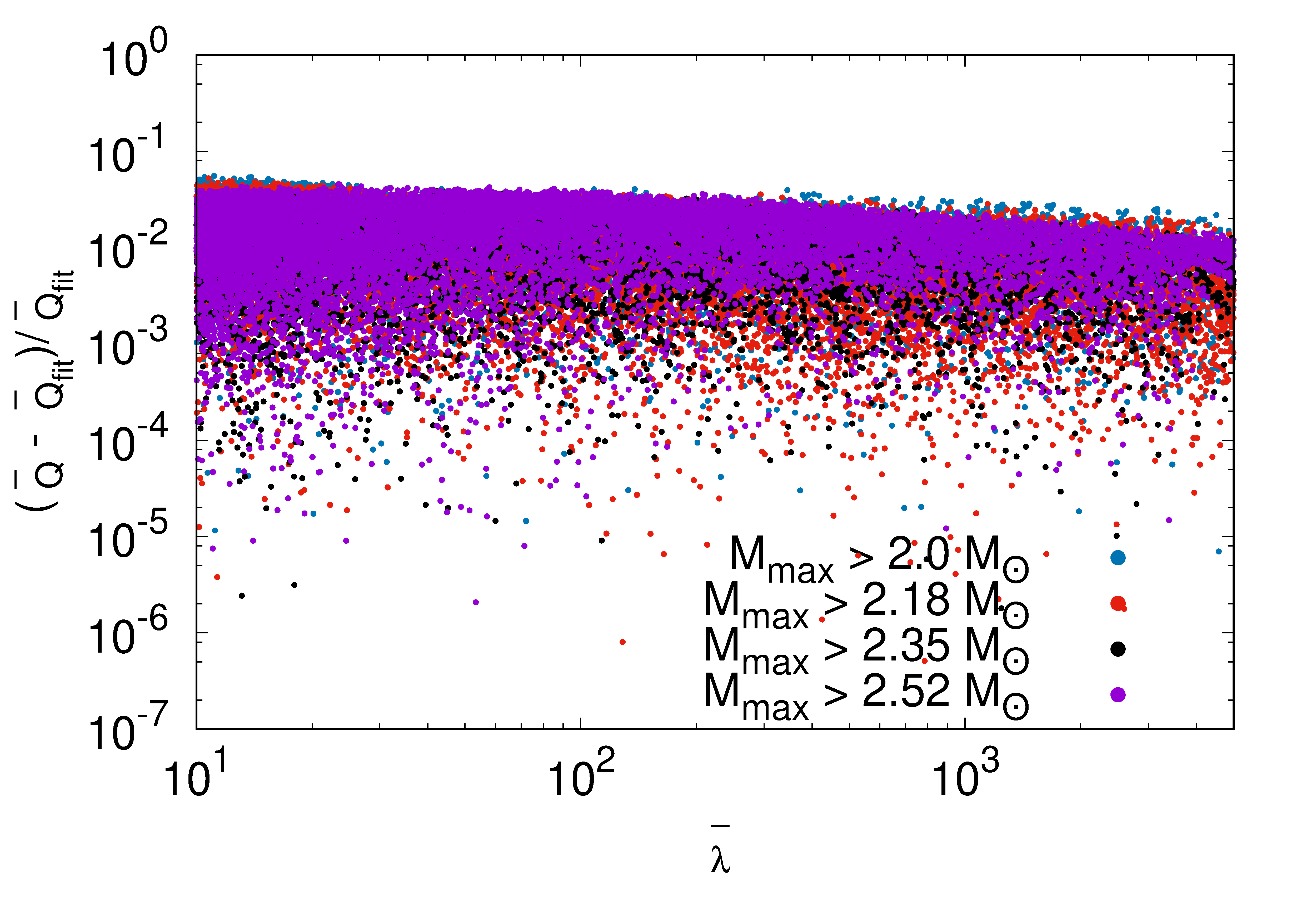}
	\caption{The fractional shift in the values of  Quadrupole moment of NSs from the fitting function $\bar{Q}_{fit}$ is shown. The nomenclature for the colour grading remains the same as fig \ref{f1}. The deviations are the same for all the maximum mass bounds.}
	\label{f8}
\end{figure}

Although the trio (I-Love-Q) follows the universal relation with great accuracy, it is useful to plot how much they deviate from the logarithmic fit. The deviation of universal relations can help characterise the universality of any other set of EoSs. The deviation of $\bar{I}$ as a function of $\lambda$ is plotted in fig \ref{f7}, and the deviation of $\bar{Q}$ as a function $\bar{\lambda}$ is shown in fig \ref{f8}. The plot shows the fractional errors between the fitted curves and the numerical results. Equation~(\ref{fit}) is a numerical fit because the data in fig \ref{f5} and \ref{f6} are themselves obtained by numerically solving the Einstein structure equations. The fitting error of $\bar{Q}$ is slightly larger than that of $\bar{I}$, although, from fig  \ref{f5} and fig \ref{f6}, it looks like $\bar{Q}$ has a smaller spread than $\bar{I}$. Most importantly, the fractional error shown in the figures \ref{f7} and \ref{f8} is the same for the EoSs with different maximum masses. As the maximum mass is a property of an EoS, this is consistent with our previous knowledge about the lack of sensitivity of universal relations with the EoSs. In this article, given a large number of EoSs, we consider 5\% percentage error as the limit for the definition of universal relations. The percentage fractional error is within 5\% for both $\bar{I}$-$\bar{\lambda}$ and $\bar{Q}$-$\bar{\lambda}$ fit. Here, we consider these fitting functions as the universal relations for the rest of the article. 


\subsection{Monotonic Sound Speed in NS}

\begin{figure}
\includegraphics[width=\columnwidth]{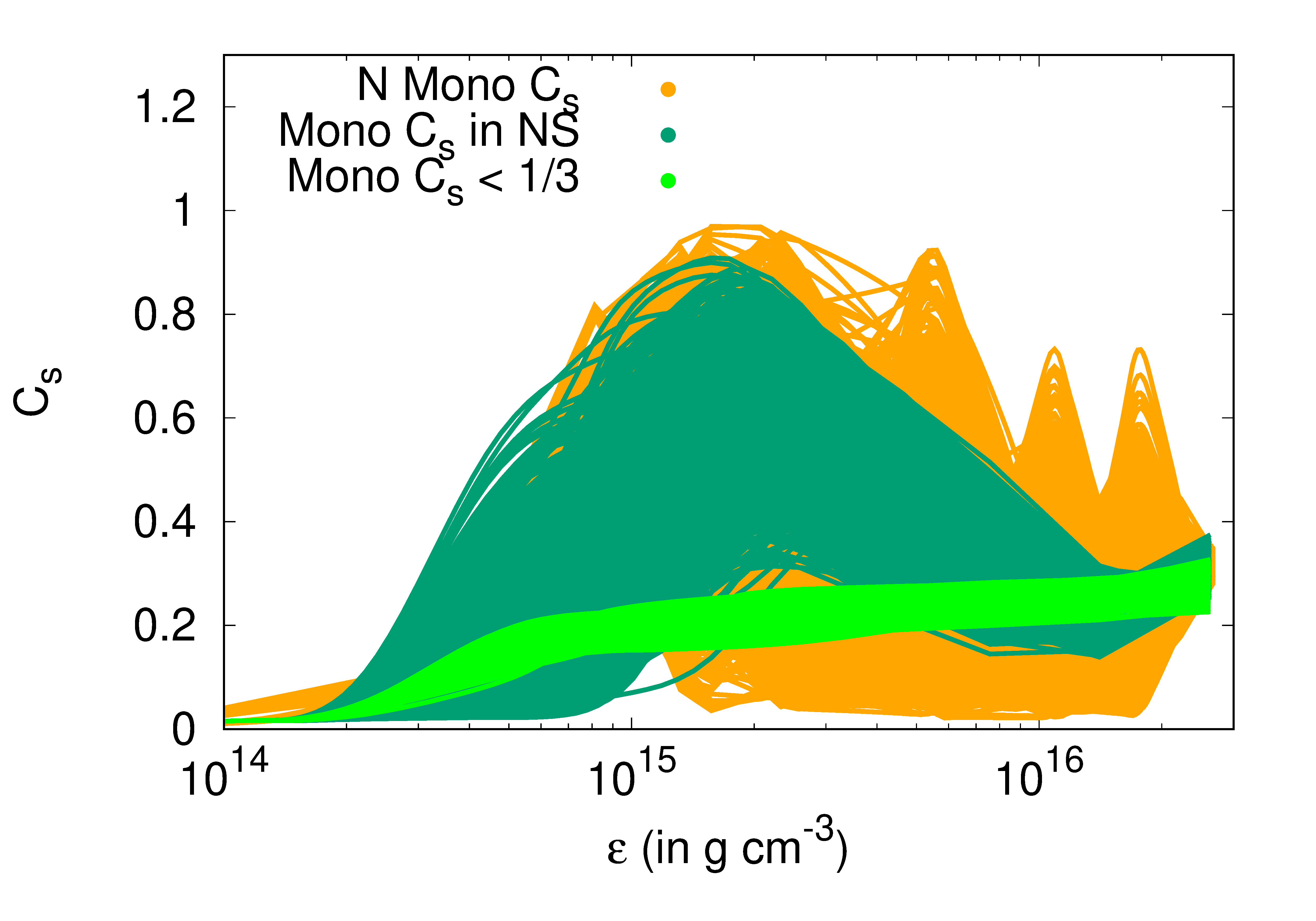}
    \caption{The speed of sound is plotted with respect to the density. The nomenclature remains the same as fig \ref{feosall}. For the dark green case, the sound speed is monotonic up to 8$\epsilon_o$. The $c_s$ remains monotonous and less than 1/3 for the bright green case. The yellow region specifies the non-monotonic sound speed that eventually exists in the EoSs which follow observational constraints.}
    \label{fcsall}
\end{figure}

\begin{figure}
\includegraphics[width=\columnwidth]{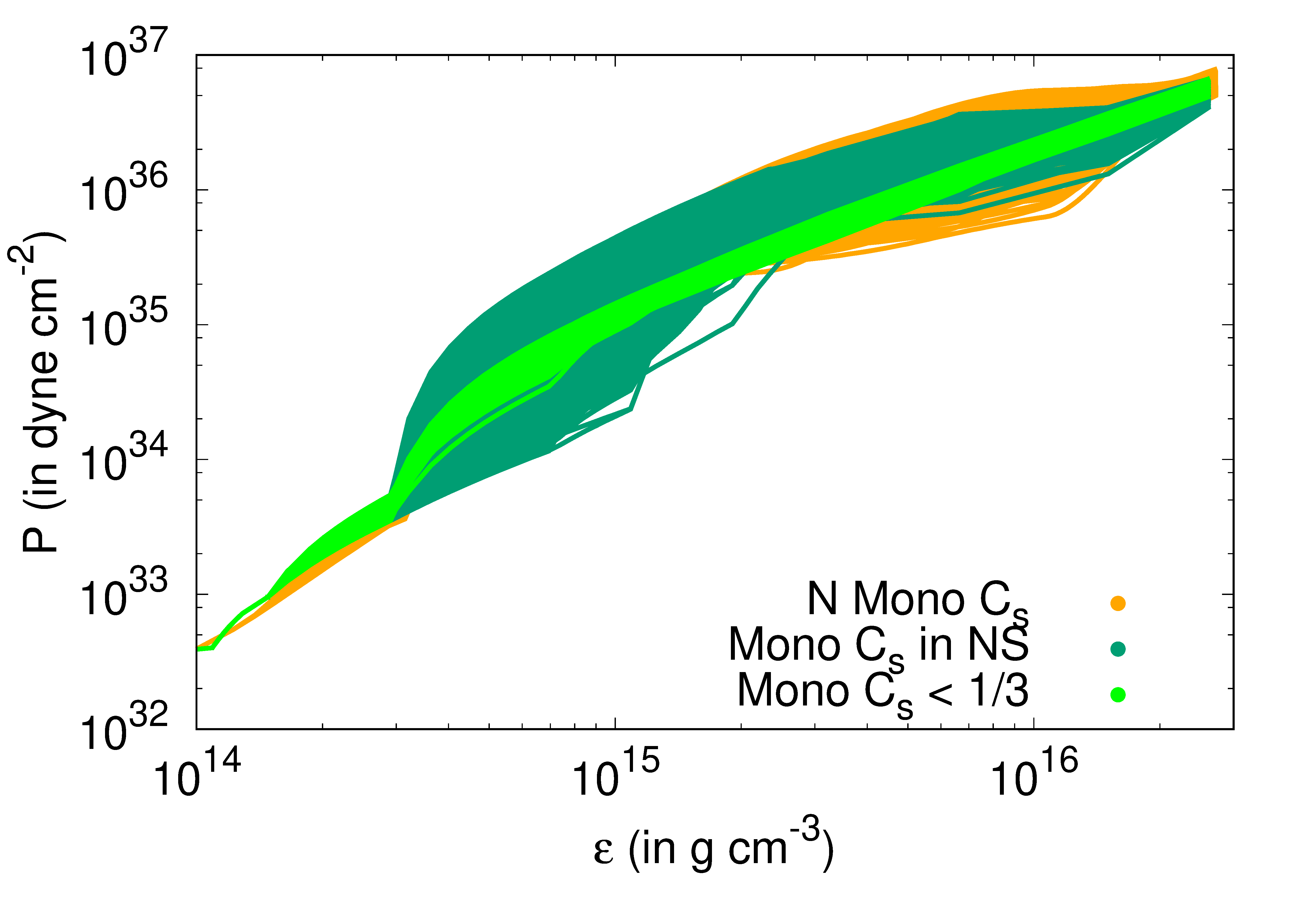}
    \caption{The generic family of EoSs that follows all astrophysical observational constraints is plotted in dark yellow. The deep green EoSs denote the EoSs with monotonic sound speed inside the star. The bright green EoSs indicate the EoSs with a monotonic sound speed of less than $1/3$. The EoSs with monotonic sound speed doesn't necessarily follow the observational constraints.}
    \label{feosall}
\end{figure}

In this section, we wanted to investigate whether universal relations are still satisfied for stars constructed with a monotonous speed of sound. These EoSs don't necessarily follow the observational constraints. There can be two cases with the monotonous speed of sound. The first case: The speed of sound remains monotonous only in the interior region of NS and reaches the limiting value of $1/3$ at asymptotically high densities. We consider that $8$ times the saturation energy density ($\approx 8 \epsilon_0$) as the maximum limiting density that the most massive star can achieve. Therefore from very low density to $8 \epsilon_0$, the sound speed is monotonous; however, after $8 \epsilon_0$, there can be non-monotonicity. \\
The second case: The speed of sound increases gradually from a very low value, monotonically increasing with density and reaching the limit of $1/3$ at asymptotically high densities; however, the speed of sound stays below $1/3$ throughout. We have constructed 2000 EoSs for both scenarios. Fig \ref{fcsall} shows how the sound velocity varies for the two cases. In addition, we have also plotted the non-monotonous sound speed that was used in the earlier section for comparison. The corresponding EoS regime is shown in fig \ref{feosall}. As the formalism for constructing EoS is different, the EoS regime differs from the non-monotonous EoS regime. However, even for these EoS, we have ensured that these are able to produce $1.4$ $M_{\odot}$ NSs.

\begin{figure}
	\includegraphics[width=\columnwidth]{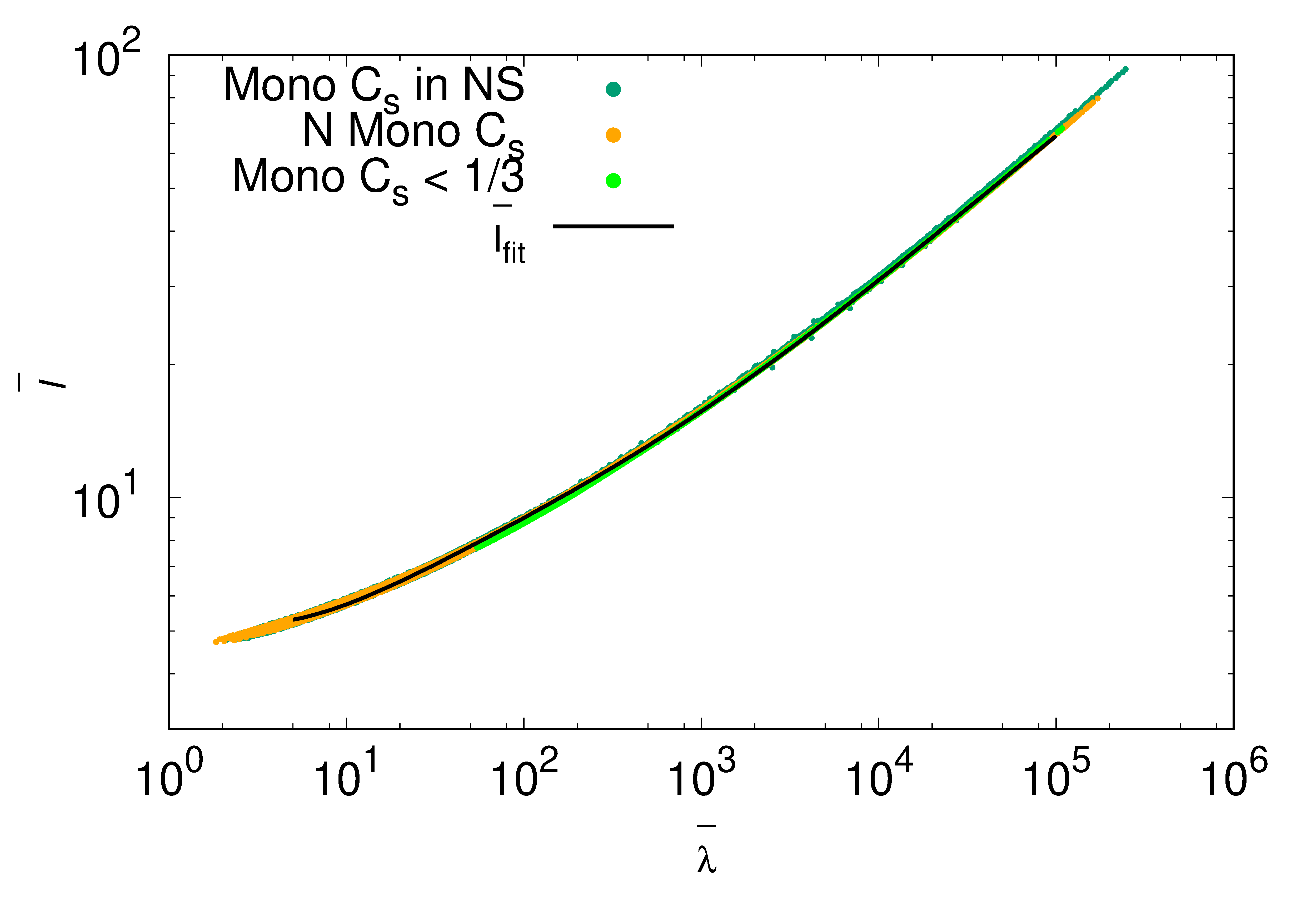}
	\caption{Moment of inertia of the NSs is plotted with respect to their tidal love number. The nomenclature for the colour grading remains the same as fig \ref{feosall}. The black line denotes the fitting function. The parameters of the fitting function are mentioned in table \ref{fit-tab}.}
	\label{f9}
\end{figure}

\begin{figure}
	\includegraphics[width=\columnwidth]{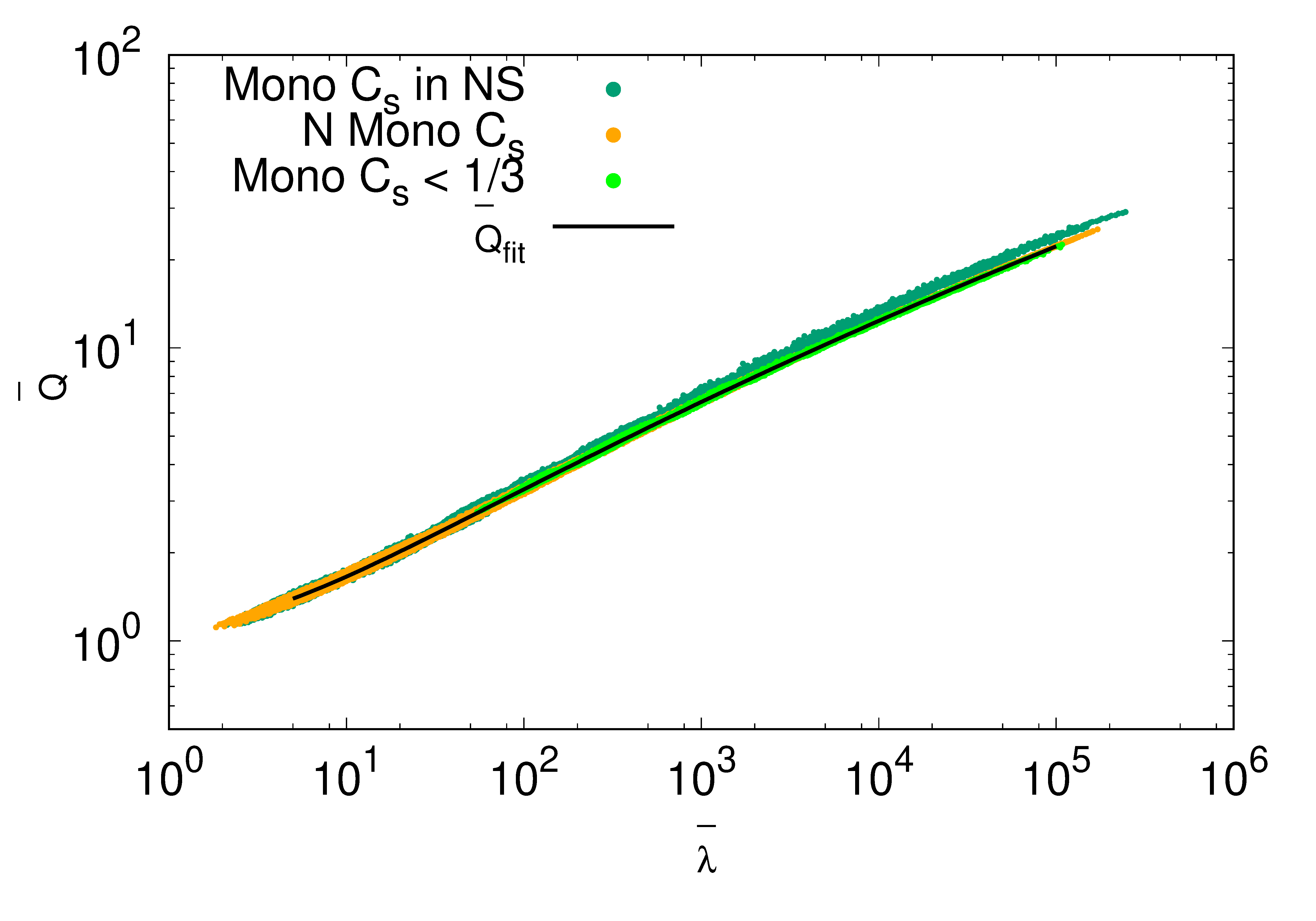}
	\caption{Quadrupole moment of the NSs is plotted with respect to their tidal love number. The nomenclature for the colour grading remains the same as fig \ref{feosall}. The black line denotes the fitting function. The parameters of the fitting function are mentioned in table \ref{fit-tab}. The EoSs with monotonic sound speed (inside the NS) seem to shift away from the fitting function.}
	\label{f10}
\end{figure}

\begin{figure}
	\includegraphics[width=\columnwidth]{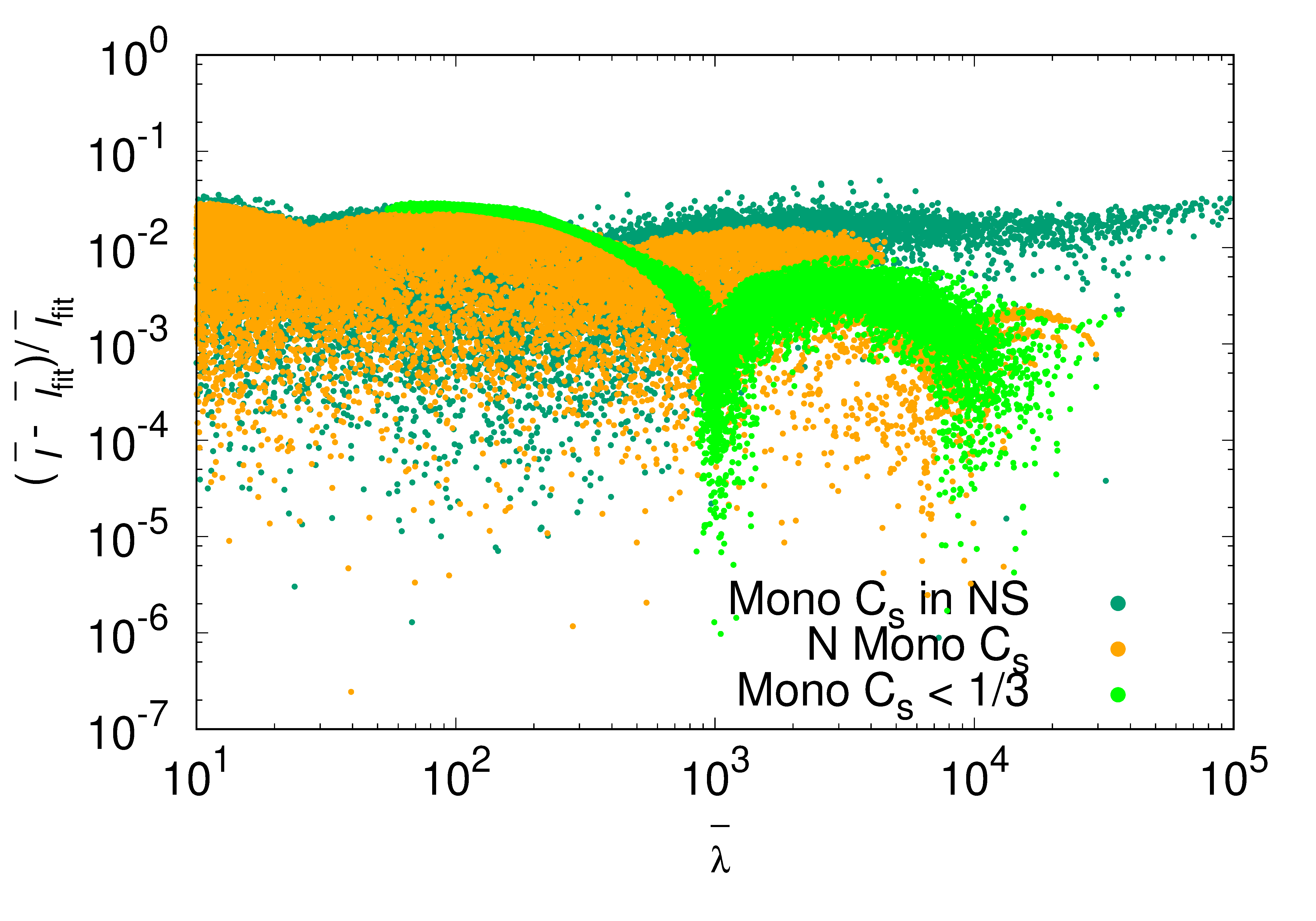}
	\caption{The deviation of the moment of inertia of NSs from the universal relation is shown. The nomenclature for the colour grading remains the same as fig \ref{feosall}. The EoSs with $c_s<1/3$ seem to produce very few stars at higher densities.}
	\label{f11}
\end{figure}

\begin{figure}
	\includegraphics[width=\columnwidth]{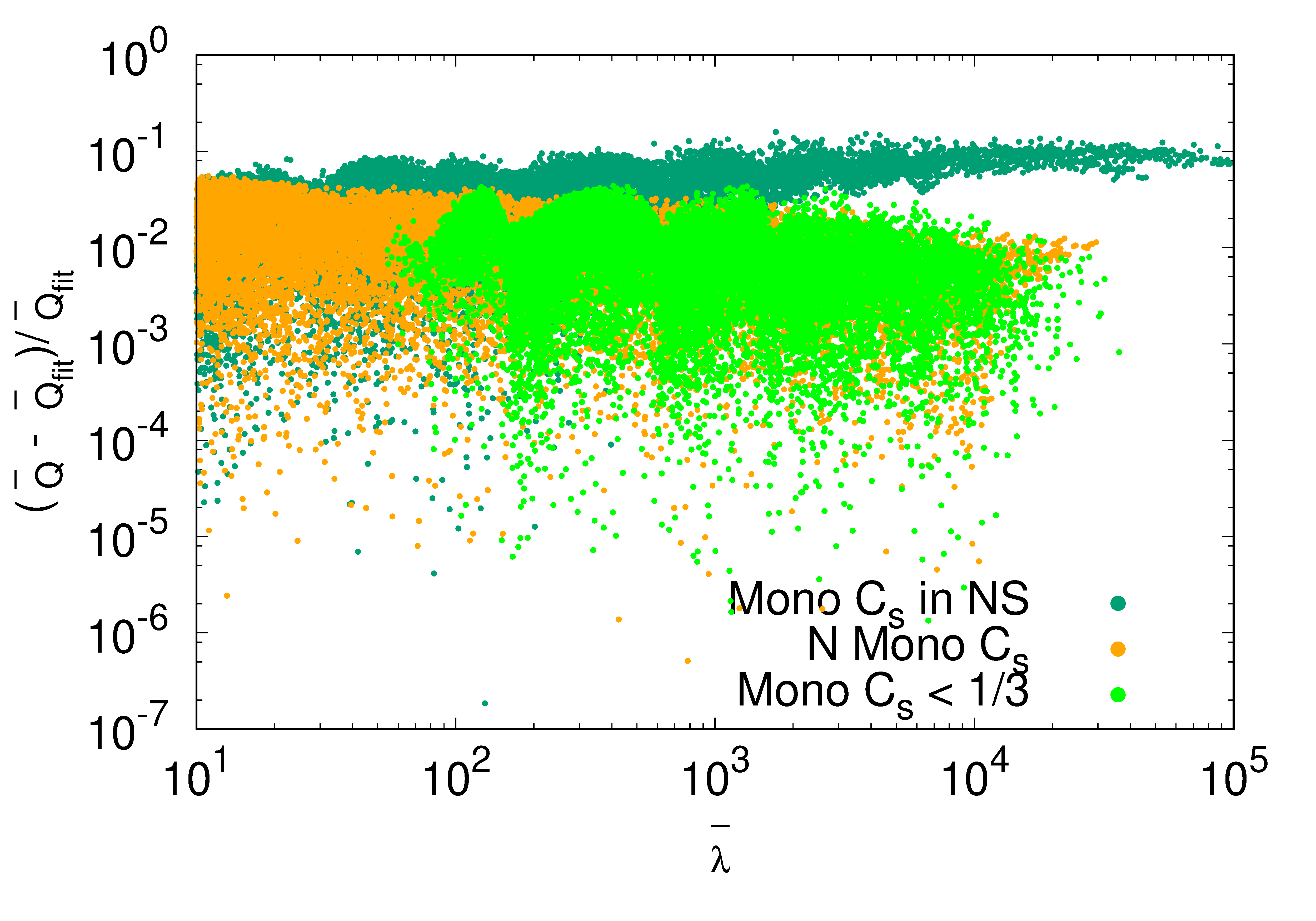}
	\caption{The deviation of the Quadrupole moment of NSs from the universal relation is shown. The nomenclature for the colour grading remains the same as fig \ref{feosall}. The fractional error reaches more than 10\% for NSs with EoSs having monotonic sound speed in the interior.}
	\label{f12}
\end{figure}

Once the EoSs are obtained, we then find their I-love-Q values. Fig \ref{f9} shows how I vary with $\bar{\lambda}$ for both monotonic and non-monotonic EoSs. Fig \ref{f10} shows how $\bar{Q}$ varies with the Love number. Both the figures are plotted for all three cases: sound speed being non-monotonic (N Mono $c_s$), monotonic inside the NS (Mono $c_s$ in NS) and monotonic sub-conformal (Mono $c_s$ < 1/3). 2000 EoSs are taken for all the 3 cases. It is seen that for the monotonic EoS, the deviation for $\bar{I}$ is much smaller. To have a clearer picture of how they deviate from the universal relation, we plot figs \ref{f11} and \ref{f12}. The deviation in the case of sub-conformal sound speed is relatively less as the EoSs are already heavily constrained. As the monotonic sound speed is below $1/3$ inside the star, this will create mostly low-mass NSs with very little spread in $\bar{Q}$ and $\bar{I}$ with respect to central density.

In fig \ref{f11}, the deviation seems to be within the limit. However, in fig \ref{f12}, we can clearly see a prominent deviation from universality. For the EoSs with monotonic sound speed inside the NS, the deviation of the Q-Love relation from universality is seen to be 10\% or more for $\bar{\lambda}$ in the range $60-800$. For the monotonic sound speed inside the NS, the deviation increases as we move towards less compact NSs. Even if the deviation is minimal for very high densities (much smaller values of $\bar{\lambda}$), the monotonic nature of $c_s$ seems to cause a shift in the matter properties as we model the NSs with typical and smaller mass. The observational constraints seem to keep the group of NSs well within the tolerance limit in the case of the generic family of NSs that have non-monotonic sound speeds. Therefore, universal relations can serve as a tool to identify EoSs that could follow observational constraints. 


\section{Conclusions}

Universal relations are important to test the theories of physics, which otherwise get obscured by the technical details. One celebrated universal relation with respect to BHs is the no-hair theorem, which determines the BH properties only by its mass and spin. Unfortunately, BHs nearest neighbours, the NSs, do not have such universal relations per se. This is because the EoS of matter at the core determines NS properties. This is the main hurdle towards it as the nature of matter at NS interior (which are at a few times nuclear saturation densities) is not clearly understood. However, similar to the no-hair theorem NSs (even QSs) have universal relations connecting the moment of inertia, quadrupole moment and tidal love number (the I-love-Q) relations. 

In the present work, we have checked the universal nature of the I-love-Q relation involving a generic family of agnostic EoS. The EoSs are generated using the bound on the speed of sound, which has to be positive but less than one. The EoS at the intermediate densities is created by randomising the speed of sound (which connects pressure and density). The lower and upper bound of the EoS is given by CET and pQCD limit, where the EoS is known with a certain degree of accuracy. We then use the constraints from recent mass and radius measurements from the NICER mission and the tidal deformability bound from GW170817. This constrains the EoS significantly. For the maximum mass bound, we have used four different maximum masses $2$, $2.18$, $2.35$ and $2.52$ solar masses. Out of $10000$ EoSs, we randomly choose $1000$ EoSs each for different maximum mass limits. We then solved the ToV equation for the static star and calculated the tidal love number for the star corresponding to each EoS. We then used the RNS code for the rotating star to obtain the moment of inertia and the quadrupole moment.

We find that both $\bar{I}$ and $\bar{Q}$ show significant spread when plotted as a function of gravitational mass. However, plotting them as a function of $\bar{\lambda}$ shows a remarkable universal nature. We have first fitted the universal nature of $\bar{I}$ vs $\bar{\lambda}$ and $\bar{Q}$ vs $\bar{\lambda}$ with a high-order logarithmic fit. Then we have also calculated the error (or spread) of the universal nature with the fitted function. We find that the deviation error for $\bar{I}$ and $\bar{Q}$ are less than 5\%. The universal relations seem to hold remarkably well. 

To check how "universal" the relations are, we have constructed two other sets of EoSs, one for which the sound speed is monotonic inside the NS and the other with sub-conformal monotonous speed throughout. This comparison is done by randomly choosing 2000 EoSs from the generic family of EoSs with non-monotonic sound speed and creating 2000 EoSs each for the two monotonic cases. A deviation of more than 10\% is seen for the NSs with EoSs with monotonic sound speed inside the star. The deviation in the case of sub-conformal sound speed is quite less as the EoSs are already heavily constrained. 

The deviation of the universal relation (from the fitting function) outside the tolerance limit is significant because it hints that such EoS differs characteristically from the original one. Even if the universal relation looks similar, one can check for deviation from the universality. If the deviation is outside a tolerance limit, one can conclude that those EoS are significantly different. The observational constraints seem to keep the group of NSs well within the tolerance limit in the case of the EoSs that have non-monotonic sound speeds. Therefore, universal relations can serve as a tool to identify EoSs that could follow observational constraints.

It should be mentioned that in most papers which use agnostic EoS, the number of EoS created is at least one order greater than what we have done in this paper. This is due to the shortcoming of a heavy computational facility in our group. However, this does not change the qualitative nature of the results that we have obtained, and even the quantitative change is expected to be negligible. Also, in the near future, we plan to have such agnostic EoS with first-order phase transition and then check the universal nature of I-love-Q.

\section*{Acknowledgements}
The authors thank IISER Bhopal for providing all the research and infrastructure facilities. SC would like to acknowledge the Prime Minister's Research Fellowship (PMRF), Ministry of Education Govt. of India, for a graduate fellowship. KKN would like to acknowledge Paolo Pani for the discussions that were quite helpful.

\section*{Data Availability}

The data used in the manuscript can be obtained on reasonable request
to the corresponding author.



\bibliographystyle{mnras}

\begin{thebibliography}{99}
\bibitem[\protect\citeauthoryear{Abbott et al.}{2018}]{Abbott} Abbott B.~P., Abbott R., Abbott T.~D., Acernese F., Ackley K., Adams C., Adams T., et al., 2018, PhRvL, 121, 161101. doi:10.1103/PhysRevLett.121.161101
\bibitem[\protect\citeauthoryear{Abbott et al.}{2019}]{Abbott1} Abbott B.~P., Abbott R., Abbott T.~D., Acernese F., Ackley K., Adams C., Adams T., et al., 2019, PhRvX, 9, 011001. doi:10.1103/PhysRevX.9.011001
\bibitem[\protect\citeauthoryear{Altiparmak, Ecker, \& Rezzolla}{2022}]{altiparmak} Altiparmak S., Ecker C., Rezzolla L., 2022, ApJL, 939, L34. 
doi:10.3847/2041-8213/ac9b2a
\bibitem[\protect\citeauthoryear{Annala et al.}{2020}]{Annala} Annala E., Gorda T., Kurkela A., N{\"a}ttil{\"a} J., Vuorinen A., 2020, NatPh, 16, 907. 
doi:10.1038/s41567-020-0914-9
\bibitem[\protect\citeauthoryear{Antoniadis et al.}{2013}]{Antoniadis} Antoniadis J., Freire P.~C.~C., Wex N., Tauris T.~M., Lynch R.~S., van Kerkwijk M.~H., Kramer M., et al., 2013, Sci, 340, 448. 
doi:10.1126/science.1233232
\bibitem[\protect\citeauthoryear{Baiotti}{2019}]{baiotti} Baiotti L., 2019, PrPNP, 109, 103714. doi:10.1016/j.ppnp.2019.103714
\bibitem[\protect\citeauthoryear{Bardeen, Carter, \& Hawking}{1973}]{hawking} Bardeen J.~M., Carter B., Hawking S.~W., 1973, CMaPh, 31, 161. 
doi:10.1007/BF01645742
\bibitem[\protect\citeauthoryear{Baym, Pethick, \& Sutherland}{1971}]{bps} Baym G., Pethick C., Sutherland P., 1971, ApJ, 170, 299. 
doi:10.1086/151216
\bibitem[\protect\citeauthoryear{Benitez et al.}{2021}]{benitez} Benitez E., Weller J., Guedes V., Chirenti C., Miller M.~C., 2021, PhRvD, 103, 023007. 
doi:10.1103/PhysRevD.103.023007
\bibitem[\protect\citeauthoryear{Bernuzzi}{2020}]{bernuzzi} Bernuzzi S., 2020, GReGr, 52, 108. doi:10.1007/s10714-020-02752-5
\bibitem[\protect\citeauthoryear{Binnington \& Poisson}{2009}]{binnington} Binnington T., Poisson E., 2009, PhRvD, 80, 084018. 
doi:10.1103/PhysRevD.80.084018
\bibitem[\protect\citeauthoryear{Chakrabarti et al.}{2014}]{chakra} Chakrabarti S., Delsate T., G{\"u}rlebeck N., Steinhoff J., 2014, PhRvL, 112, 201102. doi:10.1103/PhysRevLett.112.201102
\bibitem[\protect\citeauthoryear{Cromartie et al.}{2020}]{cromartie} Cromartie H.~T., Fonseca E., Ransom S.~M., Demorest P.~B., Arzoumanian Z., Blumer H., Brook P.~R., et al., 2020, NatAs, 4, 72. doi:10.1038/s41550-019-0880-2
\bibitem[\protect\citeauthoryear{Damour, Nagar, \& Villain}{2012}]{damour} Damour T., Nagar A., Villain L., 2012, PhRvD, 85, 123007. 
doi:10.1103/PhysRevD.85.123007
\bibitem[\protect\citeauthoryear{Doneva et al.}{2014}]{doneva} Doneva D.~D., Yazadjiev S.~S., Stergioulas N., Kokkotas K.~D., 2014, ApJL, 781, L6. 
doi:10.1088/2041-8205/781/1/L6
\bibitem[\protect\citeauthoryear{Ecker \& Rezzolla}{2022}]{ecker2} Ecker C., Rezzolla L., 2022, ApJL, 939, L35. 
doi:10.3847/2041-8213/ac8674
\bibitem[\protect\citeauthoryear{Epelbaum et al.}{2009}]{cet} Epelbaum E., Krebs H., Lee D., Mei{\ss}ner U.-G., 2009, EPJA, 40, 199. 
doi:10.1140/epja/i2009-10755-0
\bibitem[\protect\citeauthoryear{Flanagan \& Hinderer}{2008}]{flanagan} Flanagan {\'E}. {\'E}., Hinderer T., 2008, PhRvD, 77, 021502. 
doi:10.1103/PhysRevD.77.021502
\bibitem[\protect\citeauthoryear{Fraga, Kurkela, \& Vuorinen}{2014}]{fraga} Fraga E.~S., Kurkela A., Vuorinen A., 2014, ApJL, 781, L25. 
doi:10.1088/2041-8205/781/2/L25
\bibitem[\protect\citeauthoryear{Haskell et al.}{2014}]{Haskell} Haskell B., Ciolfi R., Pannarale F., Rezzolla L., 2014, MNRAS, 438, L71. 
doi:10.1093/mnrasl/slt161
\bibitem[\protect\citeauthoryear{Hebeler et al.}{2013}]{heb} Hebeler K., Lattimer J.~M., Pethick C.~J., Schwenk A., 2013, ApJ, 773, 11. 
doi:10.1088/0004-637X/773/1/11
\bibitem[\protect\citeauthoryear{Hinderer}{2008}]{hinderer} Hinderer T., 2008, ApJ, 677, 1216. 
doi:10.1086/533487
\bibitem[\protect\citeauthoryear{Kurkela et al.}{2014}]{Kurkela0} Kurkela A., Fraga E.~S., Schaffner-Bielich J., Vuorinen A., 2014, ApJ, 789, 127. 
doi:10.1088/0004-637X/789/2/127
\bibitem[\protect\citeauthoryear{Kurkela, Romatschke, \& Vuorinen}{2010}]{Kurkela1} Kurkela A., Romatschke P., Vuorinen A., 2010, PhRvD, 81, 105021. 
doi:10.1103/PhysRevD.81.105021
\bibitem[\protect\citeauthoryear{Kyutoku, Shibata, \& Taniguchi}{2021}]{Kyutoku} Kyutoku K., Shibata M., Taniguchi K., 2021, LRR, 24, 5. 
doi:10.1007/s41114-021-00033-4
\bibitem[\protect\citeauthoryear{Lackey \& Wade}{2015}]{lackey} Lackey B.~D., Wade L., 2015, PhRvD, 91, 043002. 
doi:10.1103/PhysRevD.91.043002
\bibitem[\protect\citeauthoryear{Lattimer \& Prakash}{2004}]{lat} Lattimer J.~M., Prakash M., 2004, Sci, 304, 536. doi:10.1126/science.1090720

\bibitem[\protect\citeauthoryear{Mallick, Kuzur, \& Nandi}{2022}]{debojoti} Mallick R., Kuzur D., Nandi R., 2022, EPJC, 82, 512. 
doi:10.1140/epjc/s10052-022-10468-w
\bibitem[\protect\citeauthoryear{Miller et al.}{2019}]{miller} Miller M.~C., Lamb F.~K., Dittmann A.~J., Bogdanov S., Arzoumanian Z., Gendreau K.~C., Guillot S., et al., 2019, ApJL, 887, L24. doi:10.3847/2041-8213/ab50c5
\bibitem[\protect\citeauthoryear{Miller et al.}{2021}]{miller1} Miller M.~C., Lamb F.~K., Dittmann A.~J., Bogdanov S., Arzoumanian Z., Gendreau K.~C., Guillot S., et al., 2021, ApJL, 918, L28. doi:10.3847/2041-8213/ac089b
\bibitem[\protect\citeauthoryear{Nozawa et al.}{1998}]{nozawa} Nozawa T., Stergioulas N., Gourgoulhon E., Eriguchi Y., 1998, A\&AS, 132, 431. 
doi:10.1051/aas:1998304
\bibitem[\protect\citeauthoryear{Pani, Gualtieri, \& Ferrari}{2015}]{Pani} Pani P., Gualtieri L., Ferrari V., 2015, PhRvD, 92, 124003. 
doi:10.1103/PhysRevD.92.124003
\bibitem[\protect\citeauthoryear{Pappas \& Apostolatos}{2014}]{pappas} Pappas G., Apostolatos T.~A., 2014, PhRvL, 112, 121101. 
doi:10.1103/PhysRevLett.112.121101
\bibitem[\protect\citeauthoryear{Rezzolla, Most, \& Weih}{2018}]{rezzolla} Rezzolla L., Most E.~R., Weih L.~R., 2018, ApJL, 852, L25. 
doi:10.3847/2041-8213/aaa401
\bibitem[\protect\citeauthoryear{Riley et al.}{2019}]{riley} Riley T.~E., Watts A.~L., Bogdanov S., Ray P.~S., Ludlam R.~M., Guillot S., Arzoumanian Z., et al., 2019, ApJL, 887, L21. doi:10.3847/2041-8213/ab481c
\bibitem[\protect\citeauthoryear{Riley et al.}{2021}]{riley1} Riley T.~E., Watts A.~L., Ray P.~S., Bogdanov S., Guillot S., Morsink S.~M., Bilous A.~V., et al., 2021, ApJL, 918, L27. doi:10.3847/2041-8213/ac0a81
\bibitem[\protect\citeauthoryear{Robinson}{1975}]{robinson} Robinson D.~C., 1975, PhRvL, 34, 905. 
doi:10.1103/PhysRevLett.34.905
\bibitem[\protect\citeauthoryear{Romani et al.}{2022}]{romani} Romani R.~W., Kandel D., Filippenko A.~V., Brink T.~G., Zheng W., 2022, ApJL, 934, L17. 
doi:10.3847/2041-8213/ac8007
\bibitem[\protect\citeauthoryear{Sham et al.}{2015}]{Sham} Sham Y.-H., Chan T.~K., Lin L.-M., Leung P.~T., 2015, ApJ, 798, 121. 
doi:10.1088/0004-637X/798/2/121
\bibitem[\protect\citeauthoryear{Shuryak}{1980}]{Shuryak} Shuryak E.~V., 1980, PhR, 61, 71. 
 doi:10.1016/0370-1573(80)90105-2
 \bibitem[\protect\citeauthoryear{Stergioulas \& Friedman}{1995}]{stergioulas} Stergioulas N., Friedman J.~L., 1995, ApJ, 444, 306. doi:10.1086/175605
\bibitem[\protect\citeauthoryear{Tews et al.}{2013}]{Hebeler} Tews I., Kr{\"u}ger T., Hebeler K., Schwenk A., 2013, PhRvL, 110, 032504. 
doi:10.1103/PhysRevLett.110.032504
\bibitem[\protect\citeauthoryear{Yagi \& Yunes}{2013}]{yagi} Yagi K., Yunes N., 2013, Sci, 341, 365. 
doi:10.1126/science.1236462
\bibitem[\protect\citeauthoryear{Yeung et al.}{2021}]{yeung} Yeung C.-H., Lin L.-M., Andersson N., Comer G., 2021, Univ, 7, 111. 
doi:10.3390/universe7040111




\end{thebibliography}





\bsp	
\label{lastpage}
\end{document}